\documentclass[journal]{IEEEtran}
\ifCLASSINFOpdf
\else
\fi
\usepackage{mwe} % to get dummy images
\usepackage{amsmath}
\usepackage{amssymb}
\usepackage{graphicx}
\usepackage{booktabs}%hua biao ge
\usepackage{tikz}
\usepackage{algorithm}
\usepackage{algorithmic}
\usetikzlibrary{spy}
\usepackage{bm}
\usepackage[justification=centering]{caption}

\newcommand{\param}{\boldsymbol{\theta}}
\newcommand{\x}{\mathbf{x}}
\newcommand{\y}{\mathbf{y}}
\newcommand{\R}{\mathbb{R}}

\begin{document}
\title{Combining Deep Learning and Adaptive Sparse Modeling for Low-dose CT Reconstruction}

\author{Ling~Chen,
	Zhishen~Huang,
	Yong~Long*,
	Saiprasad~Ravishankar% <-this % stops a space
\thanks{L. Chen, and Y. Long are with the University of Michigan
	Shanghai Jiao Tong University Joint Institute, Shanghai Jiao Tong University,
	Shanghai 200240, China (chen$\_$ling@sjtu.edu.cn; yong.long@sjtu.edu.cn).
	
	Z. Huang is with the Department of Computational Mathematics, Science and Engineering, East Lansing, MI 48824, USA (huangz78@msu.edu).
	
	S. Ravishankar is with the Department of Computational Mathematics, Science and Engineering and the Department of Biomedical Engineering, Michigan State University, East Lansing, MI 48824, USA (ravisha3@msu.edu).
	
	*Y. Long is the corresponding author.}}% <-this % stops a space

\maketitle

\begin{abstract}
Traditional model-based image reconstruction (MBIR) methods combine forward and noise models with simple object priors. Recent application of deep learning methods for image reconstruction provides a successful data-driven approach to addressing the challenges when reconstructing images with measurement undersampling or various types of noise. In this work, we propose a hybrid supervised-unsupervised learning framework for X-ray computed tomography (CT) image reconstruction. The proposed learning formulation leverages both sparsity or unsupervised learning-based priors and neural network reconstructors to simulate a fixed-point iteration process. Each proposed trained block consists of a deterministic MBIR solver and a neural network. The information flows in parallel through these two reconstructors and is then optimally combined, and multiple such blocks are cascaded to form a reconstruction pipeline.  We demonstrate the efficacy of this learned hybrid model for low-dose CT image reconstruction with limited training data, where we use the NIH AAPM Mayo Clinic Low Dose CT Grand Challenge dataset for training and testing. In our experiments, we study combinations of supervised deep network reconstructors and sparse representations-based (unsupervised) learned or analytical priors. Our results demonstrate the promising performance of the proposed framework compared to recent reconstruction methods.
\end{abstract}

\begin{IEEEkeywords}
Low-dose X-ray CT, image reconstruction, deep
learning, transform learning, optimal combination.
\end{IEEEkeywords}

% \fi
\IEEEpeerreviewmaketitle

\section{Introduction}
\IEEEPARstart%{T}{his} demo file is intended to serve as a ``starter file''
%for IEEE journal papers produced under \LaTeX\ using
%IEEEtran.cls version 1.8b and later.
%% You must have at least 2 lines in the paragraph with the drop letter
%% (should never be an issue)
%I wish you the best of success.
X-ray computed tomography (CT) is widely used in industrial and clinical applications. 
%Owing to the physical nature of CT measurement, it is ideal 
It is highly valuable
to reduce patients' exposure to X-ray radiation during scans by reducing the dosage. However, this creates challenges for image reconstruction.  
The conventional CT image reconstruction methods include analytical methods and model-based iterative reconstruction (MBIR) methods \cite{MBIR_review_21}. The performance of analytical methods such as the filtered back-projection (FBP) \cite{1984Practical} degrades due to the greater influence of noise in the low X-ray dose setting. 
MBIR methods aim to address such performance degradation in the low-dose X-ray computed tomography (LDCT) setting.
% MBIR methods can be divided into nonadaptive methods and learning based methods.
MBIR methods often use penalized weighted least squares (PWLS) reconstruction formulations involving
simple priors for the underlying object such as 
edge-preserving (EP) regularization that assumes the image is approximately sparse in the gradient domain. 
More recent dictionary learning-based methods~\cite{xu:12:ldx} provide improved image reconstruction quality compared to nonadaptive MBIR schemes, but
involve expensive computations for sparse encoding. %despite improved image reconstruction quality. 
% Ravishankar and Bresler proposed sparsifying transform \cite{Ravishankar2013Closed} which can obtain simple closed-form solutions for sparse coding is more efficient. 
Recent PWLS methods with regularizers involving learned sparsifying transforms (PWLS-ST~\cite{2017Low}) or a union of learned transforms (PWLS-ULTRA~\cite{zheng2018}) combine both computational efficiency (cheap sparse coding in transform domain) and the representation power of learned models (transforms).
%the advantages of least-square formulation and the representation power of dictionary learning. 

% In the last few years, deep learning methods have been successfully applied to LDCT image reconstructions and have obtained significant improvement on imaging quality. Deep learning algorithms usually learn a deep learning based framework by the huge paired datasets (low-dose scans and regular dose scans). By the learnt framework,  high quality images can be obtained from low quality images for specific datasets. 
Data-driven (deep) learning approaches have also demonstrated success for LDCT image reconstruction (see~\cite{saijongjeff20} for a review). FBPConvNet \cite{2016Deep} is a convolutional neural network (CNN) scheme that refines the quality of FBP reconstructed (corrupted) CT images to match target or ground truth images. 
% Besides working in image domain, the CNN can also learn the relationship of training pairs in other specific domain. 
Another approach WavResNet \cite{kang2018deep} learns a set of filters that are used in constructing the encoder and decoder of the convolutional framelet denoiser to refine crude LDCT images. 
% \textcolor{red}{Update previous sentence. The work does not learn a wavelet transform! What about other deep learning methods?? or cite a review paper? Mention some cons?}
% Supervised methods often can achieve better performance when the training datasets and testing datasets have high similarity features. Nonetheless, Supervised methods requires large training datasets and may demonstrate poor performance when the testing datasets is quite different from the training datasets.
%Unsupervised methods typically requires smaller training datasets than supervised methods and have better generalization properties.
However, deep learning methods often require large training sets for effective learning and generalization.
Methods based on sparsifying transform learning typically require small training sets and have been shown to generalize reasonably to new data~\cite{zheng2018}. 
Hence, Ye et al.~\cite{2021Unified} proposed a unified supervised-unsupervised (referred to here as Serial SUPER) learning framework for LDCT image reconstruction that combined supervised deep learning and unsupervised transform learning (ULTRA) regularization for robust reconstruction.
%which includes a neural network trained to be a denoiser and a pre-learned PWLS-ULTRA model enforcing the data consistency. 
The framework alternates between a neural network-based denoising step and optimizing a cost function with data-fidelity, deep network and learned transform terms.

%In this work, we propose a parallel supervised-unsupervised (Parallel SUPER) framework for LDCT image reconstruction that combines the deep learning method 

In this work, we propose an alternative repeated parallel combination of deep network reconstructions and transform learning-based reconstructions (dubbed Parallel SUPER) for improved LDCT image reconstruction. We show that the adaptive transform sparsity-based image features complement deep network learned features in every layer with appropriate weights to provide better reconstructions than either the deep network or transform learning-based baselines themselves. The proposed parallel SUPER method also outperforms the recent Serial SUPER scheme in our experiments.

%Compared to the serial SUPER method where images go through each processing component in cascade, images flow through each component in one layer/block of the proposed framework in parallel. The proposed parallel SUPER method demonstrates stronger performance than the standalone FBPConvNet, the standalone PWLS-ULTRA, and serial SUPER framework.


%\subsection{Subsection Heading Here}
%Subsection text here.
%
%% needed in second column of first page if using \IEEEpubid
%%\IEEEpubidadjcol
%
%\subsubsection{Subsubsection Heading Here}
%Subsubsection text here.
\begin{figure*}
	\centering

		\includegraphics[height=3cm]{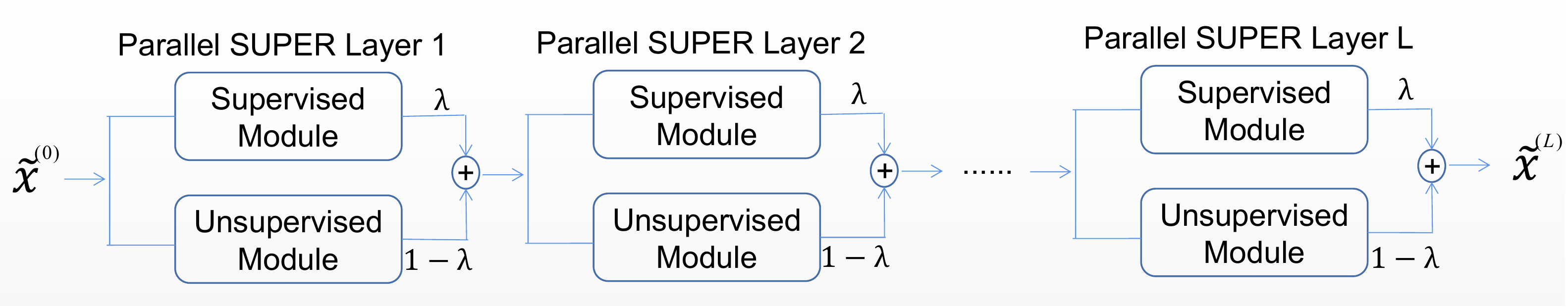}

	\vspace{-0.1in}%chuizhi jiange changdu
	\caption{Overall structure of the proposed Parallel SUPER framework.}\label{architecture}
\end{figure*}
\section{Parallel SUPER Model and the Algorithm}
\label{sec:format}
% We will discuss our Parallel SUPER framework and its properties in this section.
The proposed parallel SUPER reconstruction model is shown in Fig.~\ref{architecture}.
Each layer of parallel SUPER is comprised of a neural network and a PWLS based LDCT solver with sparsity-promoting data-adaptive regularizers. The images in the pipeline flow in parallel through these two components in a layer and are combined together (with adapted weight). The framework consists of multiple such parallel SUPER layers to ensure empirical reconstruction convergence.
% The Parallel SUPER involves combining supervised (deep network) and unsupervised (adaptive sparse model) modules over many layers. 
In this work, we have used the FBPConvNet model in the supervised module and PWLS-ULTRA as the unsupervised reconstruction module with a pre-learned union of transforms. However, specific deployed modules in the parallel SUPER framework can be replaced with other parametric models or MBIR methods. 
%Fig. 1 shows the architecture of the proposed parallel SUPER framework. 
% The architecture takes an initial image as input and processes it through multiple parallel super layers. Each layer consists of a CNN learned in a supervised module and an iterative reconstruction method in an unsupervised module in sequence. 
% While we choose the FBPConvNet as the supervised method and PWLS-ULTRA as the unsupervised method.
% Our experiments show that our proposed Parallel SUPER model can improve image quality compared with  standalone supervised method, standalone unsupervised method and Serial SUPER framework.

\subsection{Supervised Module}
The supervised modules are trained sequentially.
We set the loss function during training to be the root-mean-squared error (RMSE) to enforce alignment between the refined images and ground truth images. In the $l$-th parallel SUPER layer, the optimization problem for training the neural network is:

\begin{equation}
\label{eqn::supervised_obj}
	\mathop{\min}_{\param^{(l)}} \sum_{n=1}^{N}\| G_{\param^{(l)}}(\widetilde{\x}^{(l-1)}_n) -  \x^{\star}_n  \|_2^2,   
\end{equation}
% We denote the paired low-dose and regular-dose (reference) training images as $\{(\widetilde{\x}^{(l-1)}, \x^{*}_n)\}_{n=1}^{N}$,
where $G_{\param^{(l)}}(\cdot)$ denotes the neural network mapping in the $l$-th layer with parameters $\param^{(l)}$, $\widetilde{\x}_n^{(l-1)}$ is the $n$-th input image from $(l-1)$-th layer,  $\x^{\star}_n$ is the corresponding regular-dose (reference) image or the training label. Note that the neural networks in different layers have different parameters. 

\subsection{Unsupervised Module}
For the unsupervised module of each layer, we solve the following MBIR problem to reconstruct an image $\x \in \R^{N_p}$ from the corresponding noisy sinogram data  $\y \in \R^{N_d}$: %or union of undercomplete sparsifying transforms $\{\bm{\Omega_k}\}_{k=1}^{K}$:

\begin{equation}
\label{eqn::unsupervised_obj}
	\min_{{\x\geq 0}} J(\x,\y) := \underbrace{\frac{1}{2}\|\y-\mathbf{Ax}\|^2_{\mathbf{W}}}_{:= L(\mathbf{A}\x,\y)}  + \beta \mathcal{R}(\x),
\end{equation}
where $\mathbf{W} = \mathrm{diag}\{w_i\} \in \R^{N_d\times N_d}$ is a diagonal weighting matrix with the diagonal elements $w_i$ being the estimated inverse variance of $y_i$,  $\mathbf{A} \in \R^{N_d\times N_p}$ is the system matrix of the CT scan, $L(\mathbf{A}\x,\y)$ is the
data-fidelity term, penalty $\mathcal{R}(\x)$ is a (learning-based) regularizer, and the parameter $\beta > 0$ controls the noise and resolution trade-off. 
%In our experiments, $N_p = 262144$ and $N_d = 847872$.

In this work, we use the PWLS-ULTRA method to reconstruct an image $\x$ from noisy sinogram data  $\y$ (measurements) with a union of pre-learned transforms \ $\{\bm{\Omega}_k\}_{k=1}^{K}$. The image reconstruction is done through the following nonconvex optimization problem:
%\begin{equation}
%	\widehat{\x}^{(l)}_n(\y_n)  = \mathop{\arg\min}_{ \x_n} \sum_{k=1}^{K} \sum_{j \in \mathbf{C}_k} \frac{1}{2}\|\y_n-\mathbf{Ax}_n\|^2_{\mathbf{W}} + \bigg\{ \|\bm{\Omega}_k \mathbf{P}_j \x_n - \mathbf{z}_j \|_2^2 + \gamma^2 \|\mathbf{z}_j\|_0 \bigg\}.  
%\end{equation}
\begin{equation}
	\begin{aligned} 
		&\widehat{\x}^{(l)}(\y)  = \arg \min_{\x} \bigg\{ \frac{1}{2}\|\y-\mathbf{A}\x\|^2_{\mathbf{W}} + \\
		& \min_{\mathcal{C}_k, \mathbf{z}_j}\sum_{k=1}^{K} \sum_{j \in \mathcal{C}_k} \bigg( \|\bm{\Omega}_k \mathbf{P}_j \x - \mathbf{z}_j \|_2^2 + \gamma^2 \|\mathbf{z}_j\|_0 \bigg) \bigg\},
	\end{aligned}
\end{equation}
%where $\mathbf{W}\ = \ diag\{w_i\} \in \R^{N_d\times N_d}$ is a diagonal weighting matrix with elements being the estimated inverse variance of $y_i$,  $\mathbf{A} \in \R^{N_d\times N_p}$ is the system matrix of a CT scan.
where $\widehat{\x}^{(l)}(\y)$ is the reconstructed image by the unsupervised solver in the $l$-th layer, the operator $\mathbf{P}_j \in \R^{l \times N_p}$ extracts the $j$-th patch of $l$ voxels of image $\x$ as $\mathbf{P}_j \x$, $\mathbf{z}_j$ is the corresponding sparse encoding of the image patch under a matched transform, and $\mathcal{C}_k$ denotes the indices of patches grouped into the $k$-th cluster with transform $\bm{\Omega}_k$. Minimization over $\mathcal{C}_k$ indicates the computation of the cluster assignment of each patch.
%$\bf{\Omega}$. 
The regularizer $\mathcal{R}$ includes an encoding error term and an $\ell_0$ sparsity penalty counting the number of non-zero entries with weight $\gamma^2$. The sparse encoding and clustering are computed simultaneously.
We apply the alternating minimization method from~\cite{zheng2018} (with inner iterations for updating $\x$) on the above optimization problem. The algorithm also uses a different (potentially better) initialization in each parallel SUPER layer, which may benefit solving the involved nonconvex optimization problem.

\subsection{Parallel SUPER Model}
The main idea of the Parallel SUPER framework is to combine the supervised neural networks and iterative model-based reconstruction solvers in each layer. Define $\mathcal{M}(\widetilde{\x}^{(l-1)},\y; \Gamma)$ as an iterative MBIR solver with initial solution $\widetilde{\x}^{(l-1)}$, noisy sinogram data $\y$ and hyperparameter setting $\Gamma$ to solve optimization problem \eqref{eqn::unsupervised_obj}. In the $l$-th layer, the parallel SUPER model is formulated as:

%\begin{align*}
%	&\mathop{\min}_{{\lambda}^{(l)},\theta^{(l)}} \sum_{n=1}^{N}\| G_{\theta^{(l)}}(\widetilde{\x}^{(l-1)}_n) {\lambda}^{(l)} +\widehat{\x}^{(l)}_n(\y_n)(1 - {\lambda}^{(l)} )  -  \x^{*}_n  \|_2^2   \\
%	&   s.t. \  {\lambda}^{(l)}\geq 0, {\lambda}^{(l)} \leq 1,
%	\widehat{\x}^{(l)}_n(\y_n) = \mathop{\arg\min}_{ \x_n} J(\x_n,\y_n) 	\tag{P0}
%\end{align*}
%\begin{align*}
%	&	\mathop{\min}_{\lambda^{(l)}} \sum_{n=1}^{N}\| G_{\theta^{(l)}}(\widetilde{\x}^{(l-1)}_n) {\lambda}^{(l)} +\widehat{\x}^{(l)}_n(\y_n)(1 - {\lambda}^{(l)} )  -  \x^{*} _n \|_2^2   \   s.t. \  {\lambda}^{(l)}\geq 0, {\lambda}^{(l)} \leq 1, \\
%	&	\widehat{\x}^{(l)}_n(\y_n) = \mathop{\arg\min}_{ \x_n} J(\x_n,\y_n) , \theta^{(l)} = \mathop{\min}_{\theta^{(l)}} \sum_{n=1}^{N}\| G_{\theta^{(l)}}(\widetilde{\x}^{(l-1)}_n) -  \x^{*}_n  \|_2^2   	\tag{P0}
%\end{align*}
%\begin{align}
%	&	\widetilde{\x}^{(l)}_n = {\lambda} \cdot G_{\theta^{(l)}}(\widetilde{\x}^{(l-1)}_n)  + (1 - {\lambda} ) \cdot\widehat{\x}^{(l)}_n(\y_n) \nonumber\\
%	& \mathrm{s.t.} 
%	\begin{cases}	
%	\widehat{\x}^{(l)}_n(\y_n) &= \mathop{\arg\min}_{ \x_n} J(\x_n,\y_n) \,\forall \, n ,  \\
%	 \param^{(l)} &= \mathop{\min}_{\param^{(l)}} \sum_{n=1}^{N}\| G_{\param^{(l)}}(\widetilde{\x}^{(l-1)}_n) -  \x^{\star}_n  \|_2^2.  
%	\end{cases}
%	\tag{P0}
	\label{eqn::parallel_super_target}
%\end{align}
\begin{align}
	&	\widetilde{\x}^{(l)} = {\lambda} \cdot G_{\theta^{(l)}}(\widetilde{\x}^{(l-1)})  + (1 - {\lambda} ) \cdot\widehat{\x}^{(l)}(\y) \nonumber\\
	& \mathrm{s.t.} 
	\begin{cases}	
	\widehat{\x}^{(l)}(\y) &= \mathcal{M}(\widetilde{\x}^{(l-1)},\y; \Gamma) , \\
% 	\widehat{\x}^{(l)}_n(\y_n) &= \mathop{\arg\min}_{ \widetilde{\x}^{(l-1)}} J(\widetilde{\x}^{(l-1)},\y_n) \, ,  \forall \, n ,  \\
	 \param^{(l)} &= \mathop{\min}_{\param^{(l)}} \sum_{n=1}^{N}\| G_{\param^{(l)}}(\widetilde{\x}^{(l-1)}_n) -  \x^{\star}_n  \|_2^2,  
	\end{cases}
	\tag{P0}
	\label{eqn::parallel_super_target}
\end{align}
% $\widetilde{\x}^{(l)}_n$ is the output of the $l$-th parallel SUPER layer for $n$-th image,  
% $J(\cdot,\cdot)$ is the iterative model-based reconstruction cost function, $\y_n$ is the noisy sinogram data of $n$-th input image. 
where $\lambda$ is the nonnegative weight parameter for the neural network output in each layer and it is selected and fixed in all layers. Each parallel super layer can be thought of as a weighted average between a supervised denoised image and a reconstructed low-dose image from the unsupervised solver. Repeating multiple parallel SUPER layers simulates a fixed-point iteration to generate an ultimate reconstructed image.

The Parallel SUPER training algorithm based on (P0) is shown in Algorithm 1. The Parallel SUPER reconstruction algorithm is the same except that it uses the learned network weights in each layer.
%as obtaining $\widetilde{\x}^{(l)}_n$ in training.

\begin{algorithm}[t]\begin{footnotesize}
		\caption{Parallel SUPER Training Alogorithm}
		\label{alg:A}
		{\bf Input:} %算法的输入， \hspace*{0.02in}用来控制位置，同时利用 \\ 进行换行
		$N$ pairs of reconstructed low-dose images and corresponding regular-dose reference images $\{(\widetilde{\x}^{(0)}_n, \x^{\star}_n)\}_{n=1}^{N}$, low-dose sinograms $\{\y_n\}$, weights $\mathbf{W}_n$, $\forall \ n$, number of parallel SUPER layers $L$, weight of the supervised module $\lambda$,
		
		{\bf Output:}  supervised module parameters {$\{\bm{\theta}^{(l)}\}_{l=1}^L$}.
		
		\begin{algorithmic}
			%		\REQUIRE { initial image $\x^{(0)}$,  pre-learned $\{\bm{\Omega}\}$, number of iterations $I$}\vspace{-2pt}
			
			\FOR{$l=0,1,2, \dots , L$}
			\STATE\textbf{(1) Update $\widehat{\x}^{(l)}_n(\y_n)$:} using $\widetilde{\x}^{(l-1)}_n$ as initial image, use the PWLS-ULTRA method with $\mathbf{W}_n$ ~\cite{zheng2018} to obtain each $\widehat{\x}^{(l)}_n(\y_n)$.
			
			\STATE\textbf{(2) Update $\bm{\theta}^{(l)}$:} with $N$ paired images $\{(\widetilde{\x}^{(l-1)}_n, \x^{*}_n)\}_{n=1}^{N}$, train the supervised model by solving problem \eqref{eqn::supervised_obj} to obtain $\bm{\theta}^{(l)}$. 
			
			\STATE\textbf{(3) Generate the output of $l$-th layer $\widetilde{\x}^{(l)}_n$:} use formula in~\eqref{eqn::parallel_super_target} to obtain the output $\widetilde{\x}^{(l)}_n$ $\forall$ $n$.
	
			\ENDFOR\\
		\end{algorithmic}
	\end{footnotesize}
\end{algorithm}

\section{Experiments}
% We describe the experiment setup, training procedures, and experiment results in this section.

\subsection{Experiment Setup}
In our experiments, we use the Mayo Clinics dataset established for ``the 2016 NIH-AAPM-Mayo Clinic Low Dose CT Grand Challenge'' \cite{2016TU}. We choose 520 images from 6 of 10 patients in the dataset, among which 500 slices are used for training and 20 slices are used for validation. We randomly select 20 images from the remaining 4 patients for testing. We project the regular dose CT images $\x^{\star}$ to sinograms $\y$ by adding Poisson and additive Gaussian noise to them as follows: 
\begin{equation*}
	\centering
	y_{i}=- \log \left( I_0^{-1}\max\big(\textup{Poisson}\{ I_0 e^{-[\mathbf{Ax}^\star]_i}\} + \mathcal{N}\{0,\,\sigma^2\},\epsilon \big) \right), 
\end{equation*} 
where the original number of incident photons per ray is $I_0=10^4$, the Gaussian noise variance is $\sigma^2=25$, and $\epsilon$ \cite{2018SPULTRA} is a small positive number to avoid negative measurement data when taking the logarithm.

We use {the Michigan Image Reconstruction Toolbox to construct} fan-beam CT geometry with 736 detectors~$\times$~1152 regularly spaced projection views, and a no-scatter mono-energetic source. 
%\BLUE{The scanning parameters correspond to the dataset used Siemens SOMATOM Definition CT scanners: 
{The width of each detector column is 1.2858~mm, the source to detector distance is 1085.6~mm, and the source to rotation center distance is 595~mm. We reconstruct}
images of size $512\times 512$ with the pixel size being 0.69~mm $\times$ 0.69~mm.
%$0.9766\times 0.9766~$mm$^2$.

\subsection{Parameter Settings}
In the parallel SUPER model, we use FBPConvNet as the supervised module and PWLS-ULTRA as the unsupervised module. It takes about 10 hours for training the model for 10 layers in a GTX Titan GPU graphics processor. We train models for different values of the parameter $\lambda$ (to then select an optimal value), including 0.1, 0.3, 0.5, 0.7, and 0.9.  During the training of the supervised method, we ran $4$ epochs (kept small to reduce overfitting risks) of the stochastic gradient descent (SGD) optimizer for the FBPConvNet module in each parallel SUPER layer. The training hyperparameters of FBPConvNet are set as follows: the learning rate decreases logarithmically from 0.001 to 0.0001; the batchsize is 1; and the momentum parameter is 0.99. The filters are initialized in the various networks during training with i.i.d.\ random Gaussian entries with zero mean and variance 0.005. For the unsupervised module,  we have trained a union of $5$ sparsifying transforms using $12$ slices of regular-dose CT images (which are included in the $500$ training slices). Then, we use the pre-learned union of $5$ sparsifying transforms to reconstruct images with $5$ outer iterations and $5$ inner iterations of PWLS-ULTRA. In the training and reconstruction with ULTRA, we set the parameters  $\beta=5\times 10^{3}$ and $\gamma=20$. PWLS-EP reconstruction is used as the initialization $\widetilde{\x}^{(0)}$ of the input of network in the first layer.

We compare the proposed parallel SUPER model with the unsupervised method (PWLS-EP), standalone supervised module (FBPConvNet), standalone unsupervised module (PWLS-ULTRA), and the serial SUPER model. PWLS-EP is a penalized weighted-least squares reconstruction method with edge-preserving hyperbola regularization. For the unsupervised method (PWLS-EP), we set the parameters $\delta = 20$ and $\beta = 2 ^ {15}$ and run 100 iterations to obtain convergent results. In the training of the standalone supervised module (FBPConvNet), we run 100 epochs of training to sufficiently learn the image features with low overfitting risks. In the standalone unsupervised module (PWLS-ULTRA), we use the pre-learned union of 5 sparsifying transforms to reconstruct images. We set the parameters $\beta= 10^{4}$ and $\gamma=25$, and run $1000$ alternations with 5 inner iterations to ensure good performance.  In the serial SUPER model, we run $4$ epochs of training when learning the supervised modules (FBPConvNet), and we use the pre-learned union of $5$ sparsifying transforms and set the parameters  $\beta=5\times 10^{3}$, $\gamma=20$ and $\mu = 5 \times 10^{5}$ to reconstruct images with $20$ alternations and $5$ inner iterations for the unsupervised module (PWLS-ULTRA).

\subsection{Results}
\begin{figure}[h]	
	\centering
	\begin{tabular}{c}		
	\includegraphics[width=5cm]{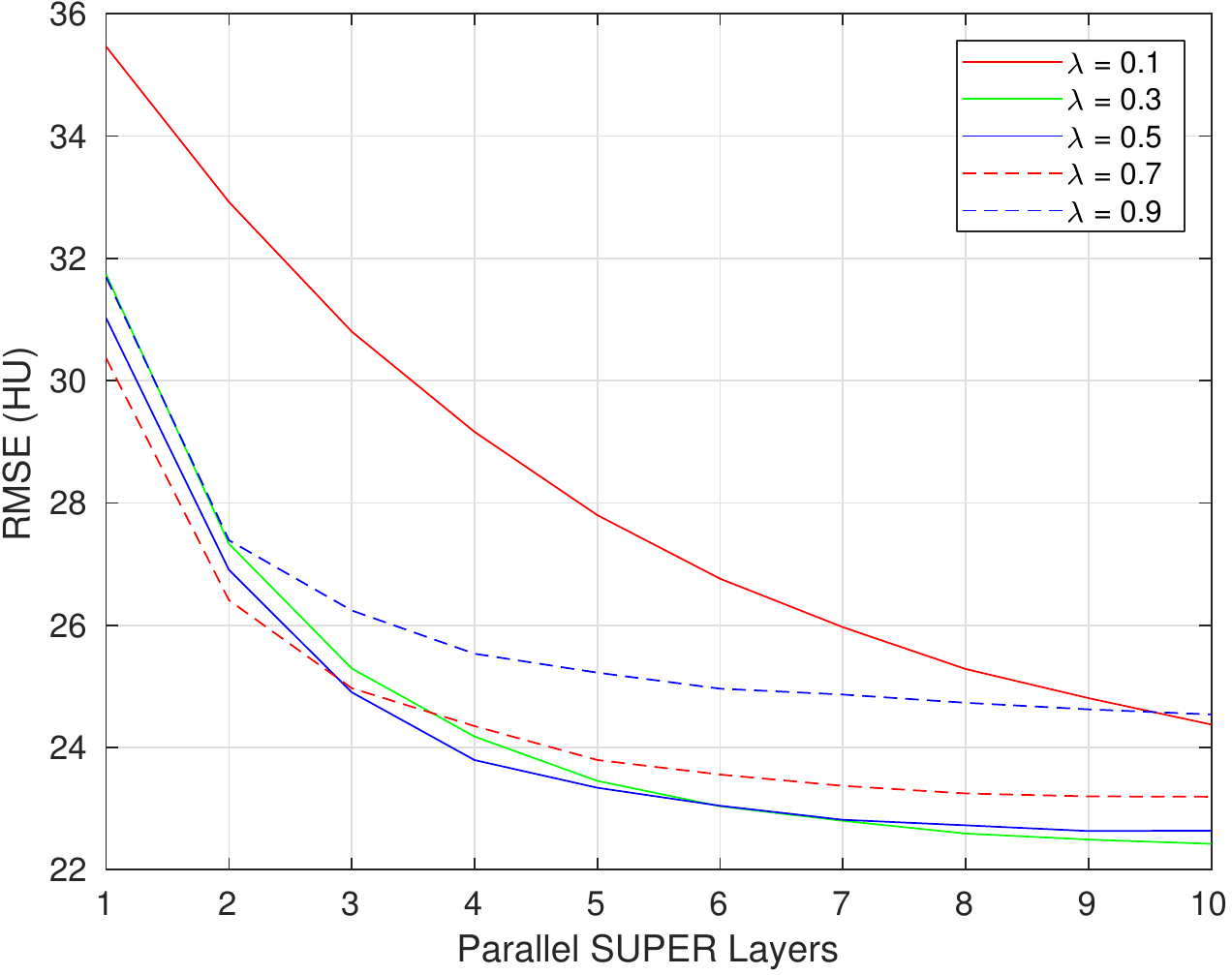}\\	\end{tabular}
	\vspace{-0.1in}%chuizhi jiange changdu
	\caption{RMSE (over 20 test slices) comparison with different choices of parameters.}\label{lambdacomparision}
\end{figure}
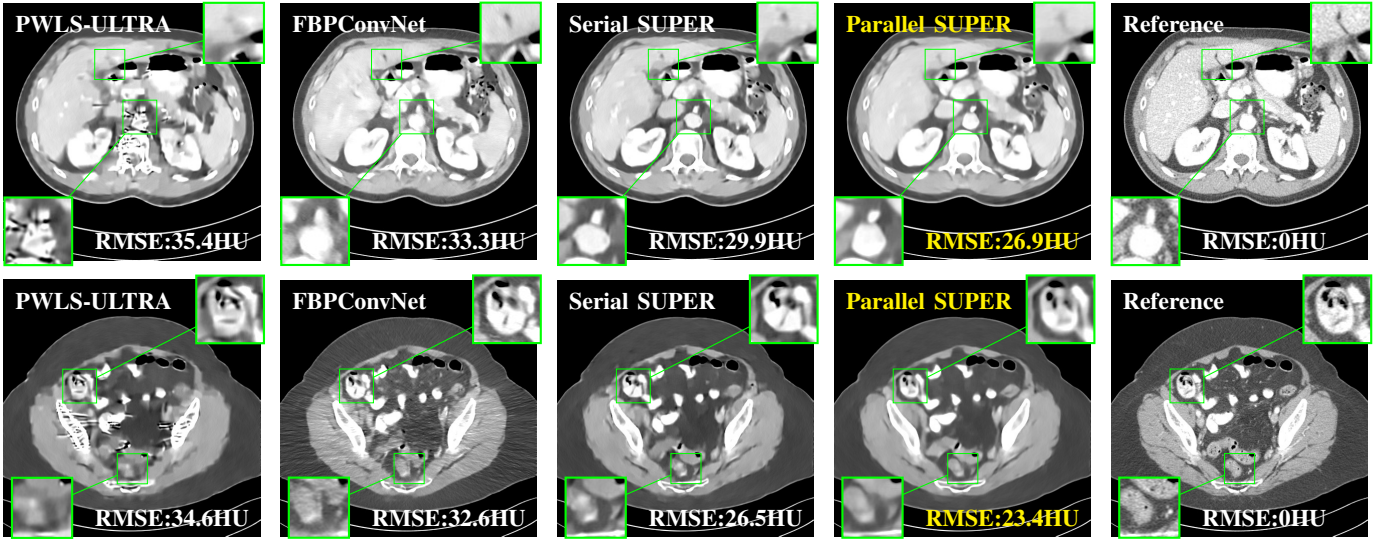
\begin{figure*}[!t]
	%	\vspace{-0.15in}
	\centering  
	\begin{tikzpicture}
		[spy using outlines={rectangle,green,magnification=2,size=9mm, connect spies}]
		\node {\includegraphics[width=0.19\textwidth]{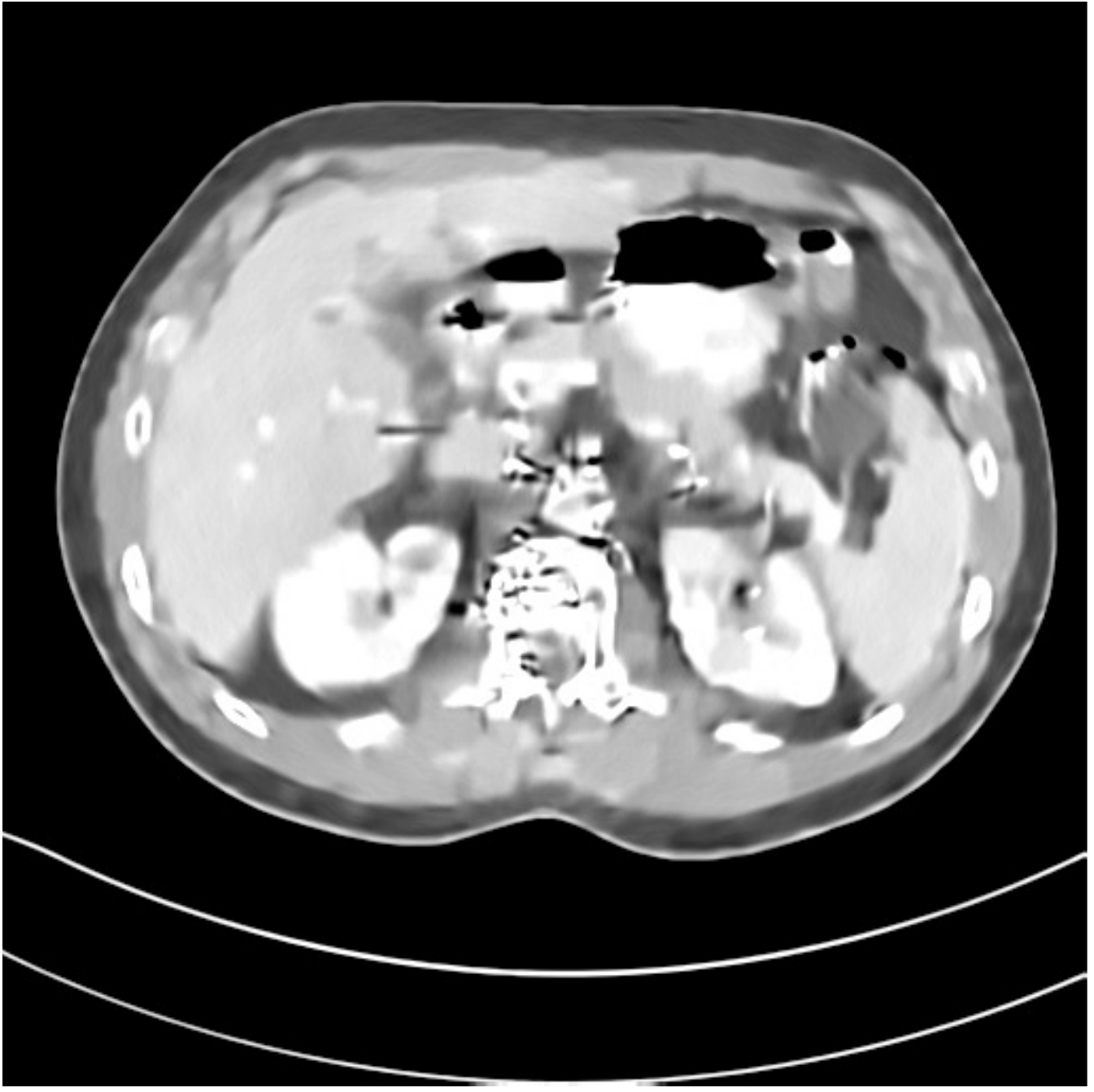} };
				\spy [green, draw, height = 0.8cm, width = 0.8cm, magnification = 2,
connect spies] on (-0.3,0.90) in node [left] at (1.75,1.32);
\spy [green, draw, height = 0.9cm, width = 0.9cm, magnification = 2,
connect spies] on (0.1,0.2) in node [left] at (-0.80,-1.32); 
		%		\spy [green, draw, height = 0.9cm, width = 0.9cm, magnification = 1.5,
		%		connect spies] on (-0.95,-0.50) in node [left] at (-0.9,-1.32); 
		%\draw[red] (-1.95,0)--(1.95,0);
		%\draw[blue] (0,-1.95)--(0,1.95);
	\end{tikzpicture}
	%	\put(-89,110){ \color{white}{\bf \small{RMSE:0.00}}}
	\put(-70,8){ \color{white}{\bf \small{RMSE:35.4HU}}}
	\put(-100,90){ \color{white}{\bf \small{PWLS-ULTRA}}} 	
	\begin{tikzpicture}
		[spy using outlines={rectangle,green,magnification=2,size=9mm, connect spies}]
		\node {\includegraphics[width=0.19\textwidth]{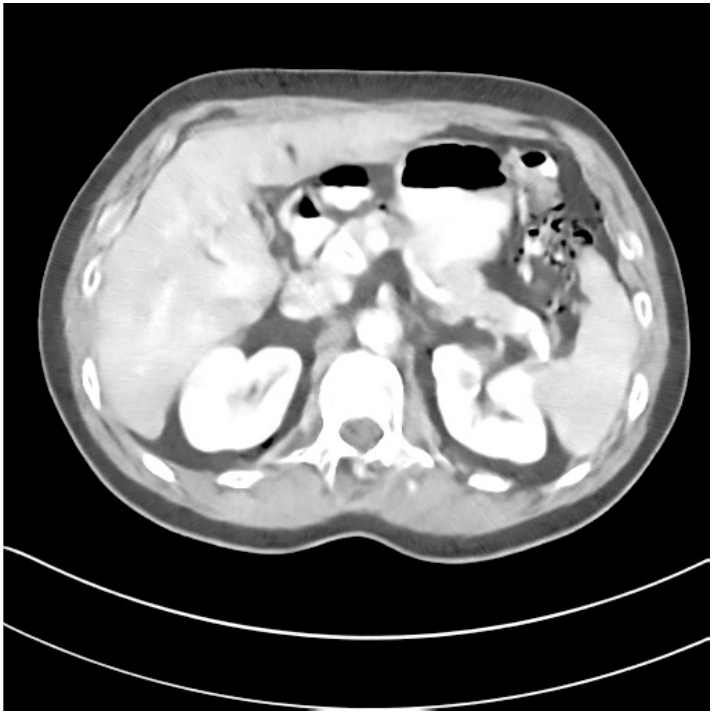} };
				\spy [green, draw, height = 0.8cm, width = 0.8cm, magnification = 2,
connect spies] on (-0.3,0.90) in node [left] at (1.75,1.32);
\spy [green, draw, height = 0.9cm, width = 0.9cm, magnification = 2,
connect spies] on (0.1,0.2) in node [left] at (-0.80,-1.32);  
		%		\spy [green, draw, height = 0.9cm, width = 0.9cm, magnification = 1.5,
		%		connect spies] on (-0.95,-0.50) in node [left] at (-0.9,-1.32); 
		%\draw[red] (-1.95,0)--(1.95,0);
		%\draw[blue] (0,-1.95)--(0,1.95);
	\end{tikzpicture}
	%	\put(-89,110){ \color{white}{\bf \small{RMSE:0.00}}}
	%	\put(-89,100){ \color{white}{\bf \small{SSIM:1.000}}}
	\put(-70,8){ \color{white}{\bf \small{RMSE:33.3HU}}}
	\put(-100,90){ \color{white}{\bf \small{FBPConvNet}}} 	
	%	\put(-80,85){ \color{white}{\bf \small{$(128\times64)$}}}
	\begin{tikzpicture}
		[spy using outlines={rectangle,green,magnification=2,size=9mm, connect spies}]
		\node {\includegraphics[width=0.19\textwidth]{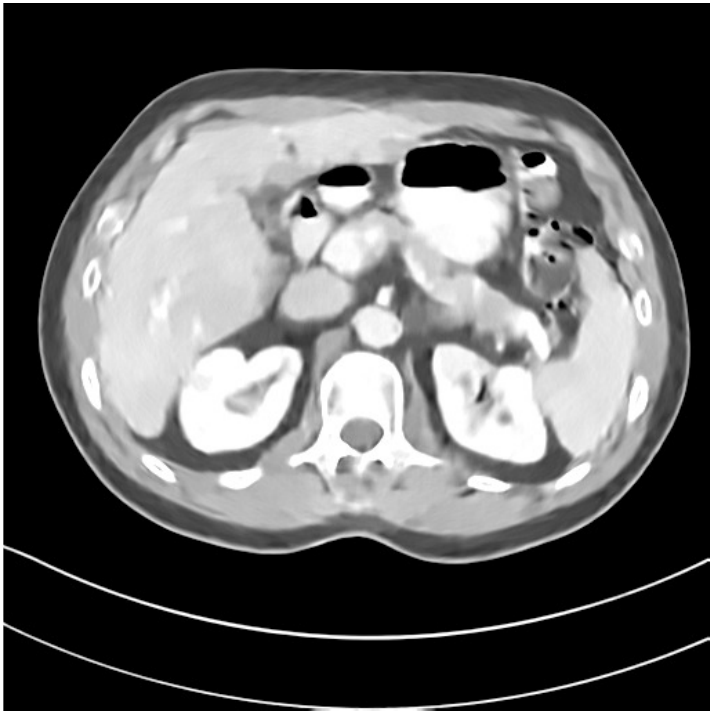}};
		
				\spy [green, draw, height = 0.8cm, width = 0.8cm, magnification = 2,
connect spies] on (-0.3,0.90) in node [left] at (1.75,1.32);
\spy [green, draw, height = 0.9cm, width = 0.9cm, magnification = 2,
connect spies] on (0.1,0.2) in node [left] at (-0.80,-1.32);  
		%		\spy [green, draw, height = 0.9cm, width = 0.9cm, magnification = 1.5,
		%		connect spies] on (-0.95,-0.50) in node [left] at (-0.9,-1.32); 
		%\draw[red] (-1.95,0)--(1.95,0);
		%\draw[blue] (0,-1.95)--(0,1.95);
	\end{tikzpicture}
	%	\put(-89,110){ \color{white}{\bf \small{RMSE:0.00}}}
	%	\put(-89,100){ \color{white}{\bf \small{SSIM:1.000}}}
	\put(-70,8){ \color{white}{\bf \small{RMSE:29.9HU}}}
	\put(-100,90){ \color{white}{\bf \small{Serial SUPER}}} 
	\begin{tikzpicture}
		[spy using outlines={rectangle,green,magnification=2,size=9mm, connect spies}]
		\node {\includegraphics[width=0.19\textwidth]{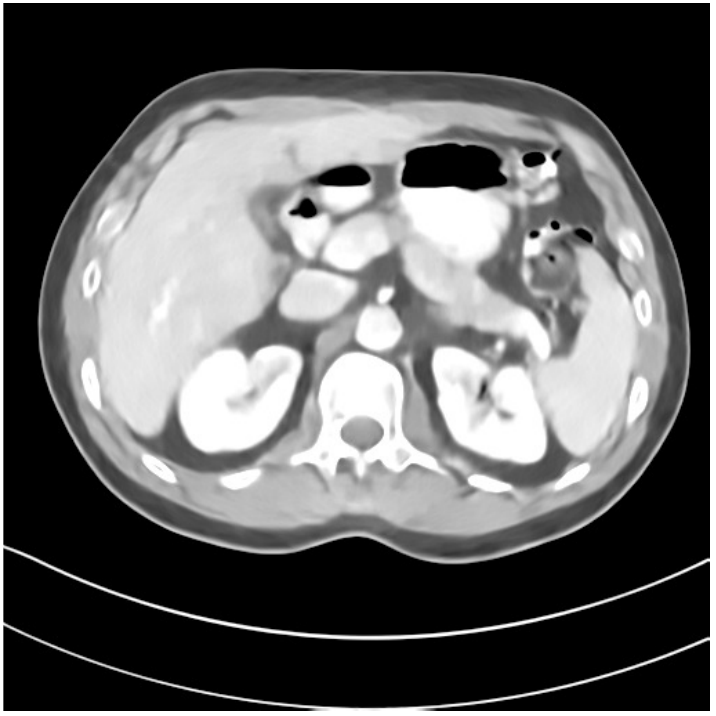} };
		
				\spy [green, draw, height = 0.8cm, width = 0.8cm, magnification = 2,
connect spies] on (-0.3,0.90) in node [left] at (1.75,1.32);
\spy [green, draw, height = 0.9cm, width = 0.9cm, magnification = 2,
connect spies] on (0.1,0.2) in node [left] at (-0.80,-1.32); 
		%		\spy [green, draw, height = 0.9cm, width = 0.9cm, magnification = 1.5,
		%		connect spies] on (-0.95,-0.50) in node [left] at (-0.9,-1.32); 
		%\draw[red] (-1.95,0)--(1.95,0);
		%\draw[blue] (0,-1.95)--(0,1.95);
	\end{tikzpicture}
	%	\put(-89,110){ \color{white}{\bf \small{RMSE:0.00}}}
	%	\put(-89,100){ \color{white}{\bf \small{SSIM:1.000}}}
	\put(-70,8){ \color{yellow}{\bf \small{RMSE:26.9HU}}}
	\put(-100,90){ \color{yellow}{\bf \small{Parallel SUPER}}} 
	%	\put(-80,85){ \color{white}{\bf \small{$(56\times64)$}}}
	\begin{tikzpicture}
		[spy using outlines={rectangle,green,magnification=2,size=9mm, connect spies}]
		\node {\includegraphics[width=0.19\textwidth]{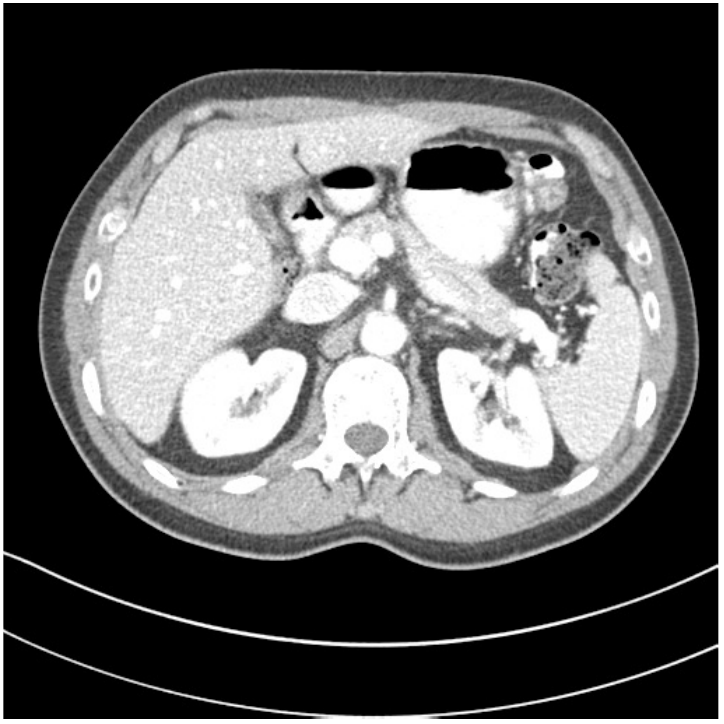} };

				\spy [green, draw, height = 0.8cm, width = 0.8cm, magnification = 2,
connect spies] on (-0.3,0.90) in node [left] at (1.75,1.32);
				\spy [green, draw, height = 0.9cm, width = 0.9cm, magnification = 2,
				connect spies] on (0.1,0.2) in node [left] at (-0.80,-1.32); 
		%		\spy [green, draw, height = 0.9cm, width = 0.9cm, magnification = 1.5,
		%		connect spies] on (-0.95,-0.50) in node [left] at (-0.9,-1.32); 
		%\draw[red] (-1.95,0)--(1.95,0);
		%\draw[blue] (0,-1.95)--(0,1.95);
	\end{tikzpicture}
	%	\put(-89,110){ \color{white}{\bf \small{RMSE:0.00}}}
	%	\put(-89,100){ \color{white}{\bf \small{SSIM:1.000}}}
	\put(-70,8){ \color{white}{\bf \small{RMSE:0HU}}}
	\put(-100,90){ \color{white}{\bf \small{Reference}}} 
	%	\put(-80,85){ \color{white}{\bf \small{$(48\times64)$}}}

		\begin{tikzpicture}
		[spy using outlines={rectangle,green,magnification=2,size=9mm, connect spies}]
		\node {\includegraphics[width=0.19\textwidth]{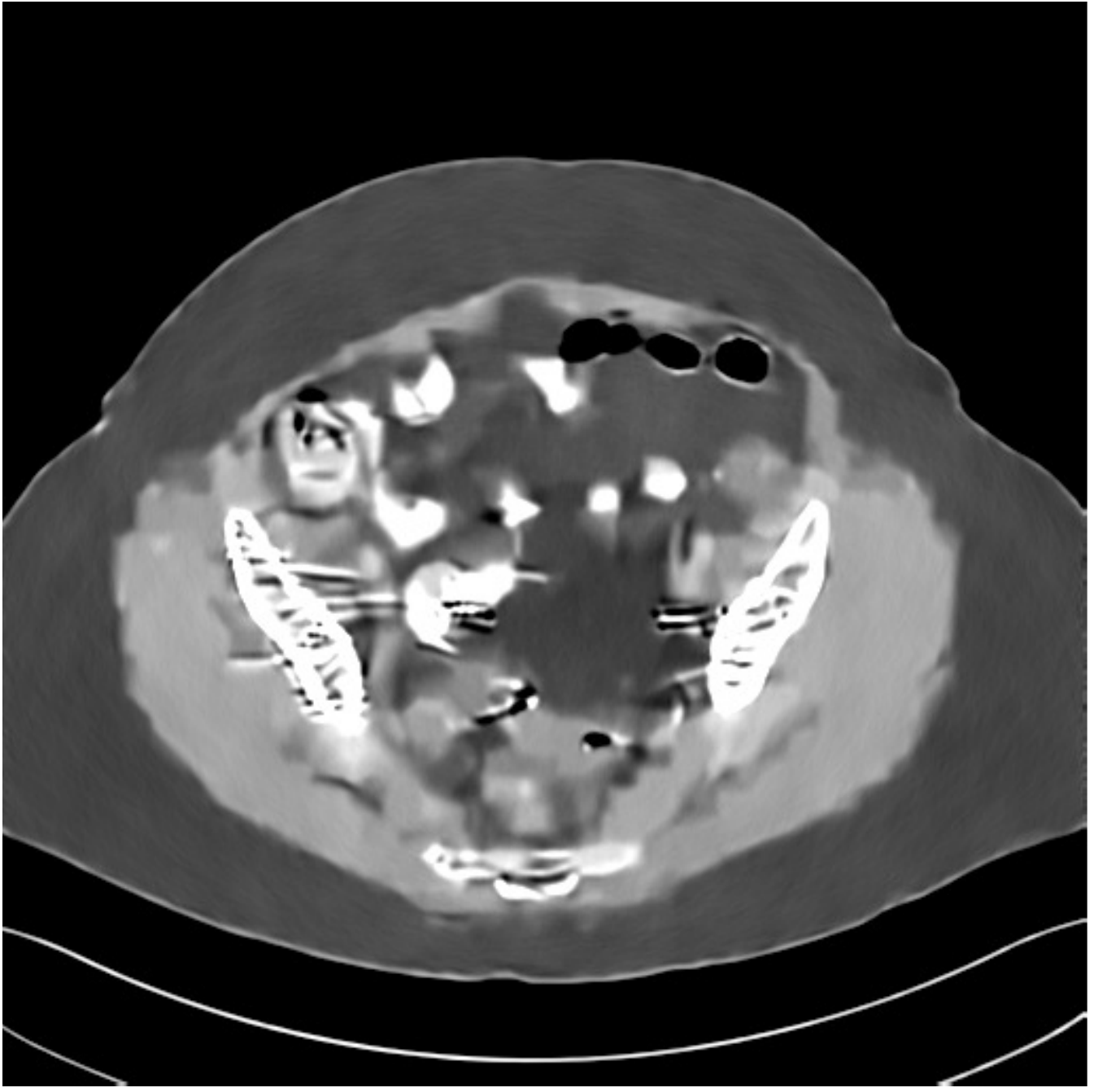} };
		
				\spy [green, draw, height = 0.8cm, width = 0.8cm, magnification = 2,
				connect spies] on (0,-0.80) in node [left] at (-0.80,-1.32);
\spy [green, draw, height = 0.9cm, width = 0.9cm, magnification = 2,
connect spies] on (-0.7,0.3) in node [left] at(1.75,1.32);
		%		\spy [green, draw, height = 0.9cm, width = 0.9cm, magnification = 1.5,
		%		connect spies] on (-0.95,-0.50) in node [left] at (-0.9,-1.32); 
		%\draw[red] (-1.95,0)--(1.95,0);
		%\draw[blue] (0,-1.95)--(0,1.95);
	\end{tikzpicture}
	%	\put(-89,110){ \color{white}{\bf \small{RMSE:0.00}}}
	\put(-70,8){ \color{white}{\bf \small{RMSE:34.6HU}}}
	\put(-100,90){ \color{white}{\bf \small{PWLS-ULTRA}}} 	
	\begin{tikzpicture}
		[spy using outlines={rectangle,green,magnification=2,size=9mm, connect spies}]
		\node {\includegraphics[width=0.19\textwidth]{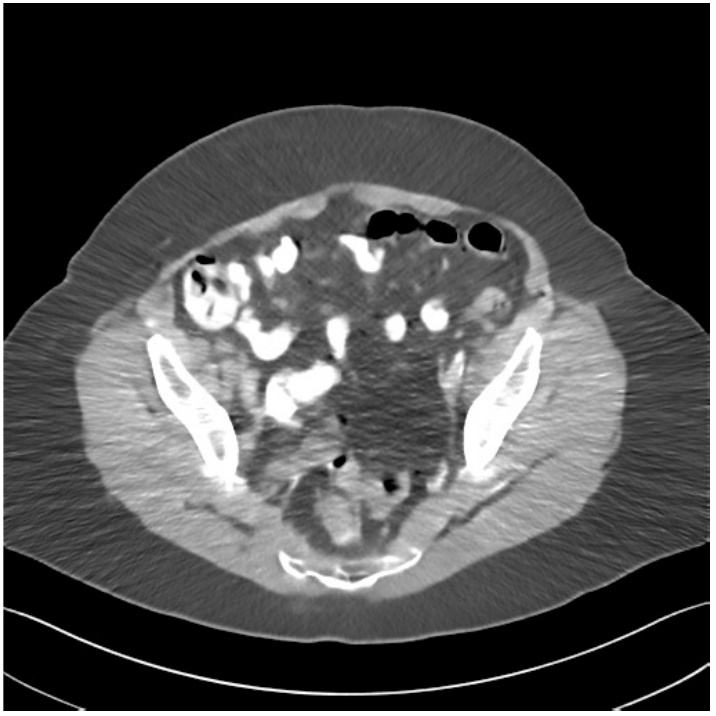} };
		
				\spy [green, draw, height = 0.8cm, width = 0.8cm, magnification = 2,
				connect spies] on (0,-0.80) in node [left] at (-0.80,-1.32);
\spy [green, draw, height = 0.9cm, width = 0.9cm, magnification = 2,
connect spies] on (-0.7,0.3) in node [left] at (1.75,1.32);
		%		\spy [green, draw, height = 0.9cm, width = 0.9cm, magnification = 1.5,
		%		connect spies] on (-0.95,-0.50) in node [left] at (-0.9,-1.32); 
		%\draw[red] (-1.95,0)--(1.95,0);
		%\draw[blue] (0,-1.95)--(0,1.95);
	\end{tikzpicture}
	%	\put(-89,110){ \color{white}{\bf \small{RMSE:0.00}}}
	%	\put(-89,100){ \color{white}{\bf \small{SSIM:1.000}}}
	\put(-70,8){ \color{white}{\bf \small{RMSE:32.6HU}}}
	\put(-100,90){ \color{white}{\bf \small{FBPConvNet}}} 	
	%	\put(-80,85){ \color{white}{\bf \small{$(128\times64)$}}}
	\begin{tikzpicture}
		[spy using outlines={rectangle,green,magnification=2,size=9mm, connect spies}]
		\node {\includegraphics[width=0.19\textwidth]{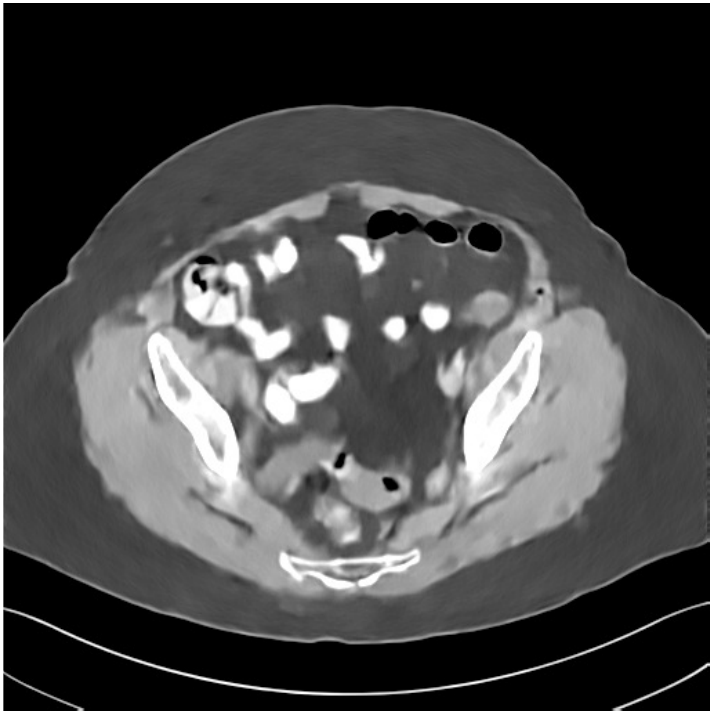}};

						\spy [green, draw, height = 0.8cm, width = 0.8cm, magnification = 2,
				connect spies] on (0,-0.80) in node [left] at (-0.80,-1.32);
\spy [green, draw, height = 0.9cm, width = 0.9cm, magnification = 2,
connect spies] on (-0.7,0.3) in node [left] at (1.75,1.32);
		%		\spy [green, draw, height = 0.9cm, width = 0.9cm, magnification = 1.5,
		%		connect spies] on (-0.95,-0.50) in node [left] at (-0.9,-1.32); 
		%\draw[red] (-1.95,0)--(1.95,0);
		%\draw[blue] (0,-1.95)--(0,1.95);
	\end{tikzpicture}
	%	\put(-89,110){ \color{white}{\bf \small{RMSE:0.00}}}
	%	\put(-89,100){ \color{white}{\bf \small{SSIM:1.000}}}
	\put(-70,8){ \color{white}{\bf \small{RMSE:26.5HU}}}
	\put(-100,90){ \color{white}{\bf \small{Serial SUPER}}} 
	\begin{tikzpicture}
		[spy using outlines={rectangle,green,magnification=2,size=9mm, connect spies}]
		\node {\includegraphics[width=0.19\textwidth]{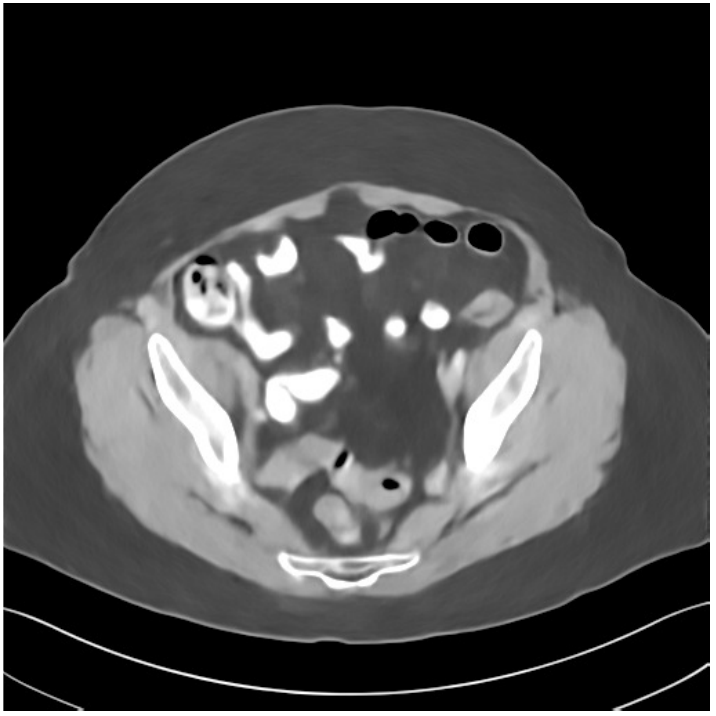} };

				\spy [green, draw, height = 0.8cm, width = 0.8cm, magnification = 2,
				connect spies] on (0,-0.80) in node [left] at (-0.80,-1.32);
\spy [green, draw, height = 0.9cm, width = 0.9cm, magnification = 2,
connect spies] on (-0.7,0.3) in node [left] at (1.75,1.32);
		%		\spy [green, draw, height = 0.9cm, width = 0.9cm, magnification = 2,
		%		connect spies] on (-0.05,-0.25) in node [left] at (0.40,-1.32); 
		%		\spy [green, draw, height = 0.9cm, width = 0.9cm, magnification = 1.5,
		%		connect spies] on (-0.95,-0.50) in node [left] at (-0.9,-1.32); 
		%\draw[red] (-1.95,0)--(1.95,0);
		%\draw[blue] (0,-1.95)--(0,1.95);
	\end{tikzpicture}
	%	\put(-89,110){ \color{white}{\bf \small{RMSE:0.00}}}
	%	\put(-89,100){ \color{white}{\bf \small{SSIM:1.000}}}
	\put(-70,8){ \color{yellow}{\bf \small{RMSE:23.4HU}}}
	\put(-100,90){ \color{yellow}{\bf \small{Parallel SUPER}}} 
	%	\put(-80,85){ \color{white}{\bf \small{$(56\times64)$}}}
	\begin{tikzpicture}
		[spy using outlines={rectangle,green,magnification=2,size=9mm, connect spies}]
		\node {\includegraphics[width=0.19\textwidth]{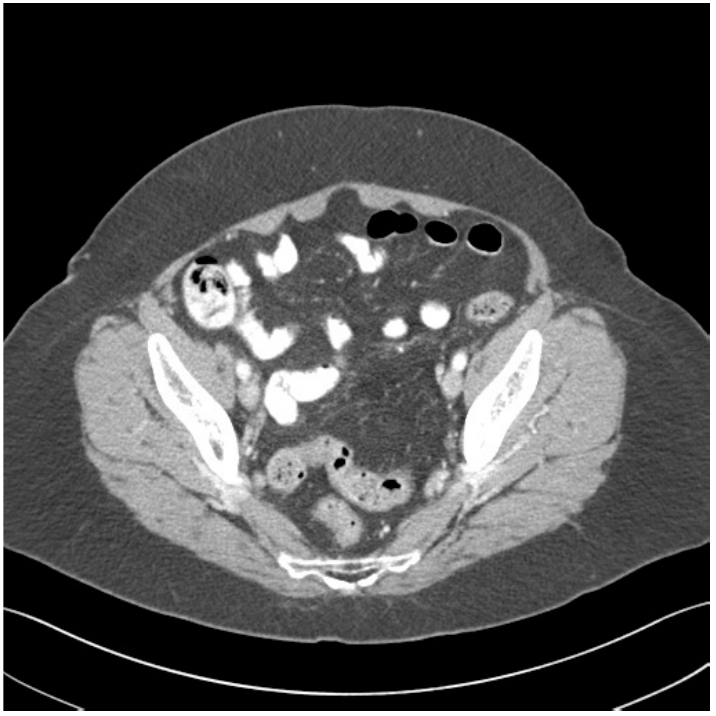} };
				\spy [green, draw, height = 0.8cm, width = 0.8cm, magnification = 2,
				connect spies] on (0,-0.80) in node [left] at (-0.80,-1.32);
				\spy [green, draw, height = 0.9cm, width = 0.9cm, magnification = 2,
connect spies] on (-0.7,0.3) in node [left] at(1.75,1.32);
		%		\spy [green, draw, height = 0.9cm, width = 0.9cm, magnification = 1.5,
		%		connect spies] on (-0.95,-0.50) in node [left] at (-0.9,-1.32); 
		%\draw[red] (-1.95,0)--(1.95,0);
		%\draw[blue] (0,-1.95)--(0,1.95);
	\end{tikzpicture}
	%	\put(-89,110){ \color{white}{\bf \small{RMSE:0.00}}}
	%	\put(-89,100){ \color{white}{\bf \small{SSIM:1.000}}}
	\put(-70,8){ \color{white}{\bf \small{RMSE:0HU}}}
	\put(-100,90){ \color{white}{\bf \small{Reference}}} 
	%	\put(-80,85){ \color{white}{\bf \small{$(48\times64)$}}}
	\vspace{-0.1in}
	\caption{Reconstruction of slice 50 from patient L067 and reconstruction of slice 150 from patient L310 using various methods. The display window is [800, 1200] HU.}\label{Resultscomparision}
	%	\vspace{-0.2in}
	%\label{fig:recon_XCAT_slice48}
\end{figure*}
%PWLS-EP is a penalized weighted-least squares reconstruction method with edge-preserving hyperbola regularization. PWLS-ULTRA is a penalized weighted-least squares reconstruction method with regularization based on a union of learnt sparsifying transforms. FBPConvNet is a CNN-based image domain denoising architecture designed for sparse-view CT. Serial SUPER is a unified framework which alternates between supervised method and unsupervised method.

To compare the performance  quantitatively,  we compute the RMSE in Hounsfield units (HU) and structural similarity index measure (SSIM) \cite{2004Image}  for the reconstructed images. For a reconstructed image $\widehat{\x}$, RMSE is defined as $\sqrt{ \sum_{j=1}^{N_{p}} (\widehat{x}_{j}-x_{j}^\star)^2/{N_{p}}}$,  where $x_{j}^\star$ denotes the reference regular-dose image intensity at the $j$-th pixel location  and $N_{p}$ is the number of pixels.
%\begin{figure}[h]	
%%	\begin{tabular}{c}		
%		%\includegraphics[height=4cm,width=4cm]{Results//UST//W56initialization.pdf}&
%		\includegraphics[width=5cm]{figure//RMSE.pdf}\\	\end{tabular}
%	\vspace{-0.1in}%chuizhi jiange changdu
%	\caption{RMSE (over test slice) comparisons with different combination parameters.}
%\end{figure}

We train the parallel SUPER framework with different choices of the parameter $\lambda$ including $0.1, 0.3, 0.5, 0.7$ and  $0.9$ to obtain the best choice. Fig.~\ref{lambdacomparision} shows the  evolution of RMSE over layers for 20 validation slices with different  $\lambda$ choices. We can see that we obtain the best RMSE when $\lambda = 0.3$.

We have conducted experiments on  20 test slices (slice 20, slice 50, slice 100, slice 150 and slice 200 of patient L067, L143, L192, L310) of the Mayo Clinic data. Table  ~\ref{Numericalresults} shows the averaged image quality of 20 test images with different methods. From Table~\ref{Numericalresults}, we observe that Parallel SUPER significantly improves the image quality compared with the standalone methods. It also achieves 1.8 HU better average RMSE compared with Serial SUPER while its SSIM is comparable with Serial SUPER.  Fig.~\ref{Resultscomparision} shows the reconstructions of L067 (slice 50) and L310 (slice 150) using PWLS-ULTRA, FBPConvNet, serial SUPER (FBPConvNet + PWLS-ULTRA), and parallel SUPER (FBPConvNet + PWLS-ULTRA), along with the references (ground truth). The Parallel SUPER scheme achieved the lowest RMSE and the zoom-in areas show that Parallel SUPER can reconstruct image details better.

\begin{table}[htbp]\normalsize
	\caption{The mean RMSE and SSIM values for 20 test images with PWLS-EP, PWLS-ULTRA, FBPConvNet, Serial SUPER, and the proposed Parallel SUPER.}
	%\vspace*{-0.1in}
	\begin{tabular}{cccc}
		\toprule
		&PWLS-EP & PWLS-ULTRA& FBPConvNet \\
		\midrule
		RMSE  & $41.4$  &$32.4$&$29.2$ \\
		\midrule
		SSIM& $0.673$&$0.716$ & $0.688$ \\
%		\midrule
%		PSNR& $0.892$&$0.965$ & $0.963$ \\
		\midrule	
		&Serial SUPER& Parallel SUPER&\\
		\midrule
		RMSE&$25.0$ & $\mathbf{23.2}$ &\\
		\midrule
		SSIM&$0.748$&  $\mathbf{0.751}$ &\\
%		\midrule
%		PSNR& $0.892$&$0.965$ & $0.963$ \\
		\bottomrule
	\end{tabular}\label{Numericalresults}
\end{table}

\section{Conclusions}
This paper proposes the parallel SUPER framework combining supervised deep learning methods and unsupervised methods for low-dose CT reconstruction. We have experimented on a setting with the supervised model FBPConvNet and the unsupervised model PWLS-ULTRA. This framework demonstrates better reconstruction accuracy and faster convergence compared to individual involved modules as well as the recent serial SUPER framework.

% or


%\appendices
%% you can choose not to have a title for an appendix
%% if you want by leaving the argument blank
%\section{}

\section*{Acknowledgment}

The authors thank Dr. Cynthia McCollough, the Mayo Clinic,
the American Association of Physicists in Medicine, and the
National Institute of Biomedical Imaging and Bioengineering for
providing the Mayo Clinic data.

\ifCLASSOPTIONcaptionsoff
  \newpage
\fi



%\bibliographystyle{IEEEtran}

\bibliographystyle{IEEEtran}

\bibliography{mybib}
%\begin{thebibliography}{1}
%\end{thebibliography}


%\end{IEEEbiography}


%\newpage



%\vfill


\end{document}